\documentclass[]{aastex631}

\usepackage{amsmath}
\usepackage{multirow}
\shorttitle{Growth of Cyclotron Waves Measured by \textit{PSP}}
\shortauthors{He et al.}
\graphicspath{{./}{figures/}}

\begin{document}

\title{Growth of Outward Propagating Fast-Magnetosonic/Whistler Waves in the Inner Heliosphere Observed by Parker Solar Probe}

\correspondingauthor{Jiansen He}
\email{jshept@pku.edu.cn}

\author[0000-0001-8179-417X]{Jiansen He}
\affiliation{School of Earth and Space Sciences, Peking University, \\
Beijing, 100871, P. R. China}

\author[0000-0001-8245-9752]{Ying Wang}
\affiliation{School of Earth and Space Sciences, Peking University, \\
Beijing, 100871, P. R. China}

\author[0000-0002-1541-6397]{Xingyu Zhu}
\affiliation{School of Earth and Space Sciences, Peking University, \\
Beijing, 100871, P. R. China}

\author[0000-0002-6300-6800]{Die Duan}
\affiliation{School of Earth and Space Sciences, Peking University, \\
Beijing, 100871, P. R. China}

\author[0000-0002-0497-1096]{Daniel Verscharen}
\affiliation{Mullard Space Science Laboratory, University College London, Dorking RH5 6NT, UK}
\affiliation{Space Science Center, University of New Hampshire, Durham NH 03824, USA}

\author[0000-0002-1831-1451]{Guoqing Zhao}
\affiliation{Institute of Space Physics, Luoyang Normal University, Luoyang, People’s Republic of China}



\begin{abstract}
The solar wind in the inner heliosphere has been observed by \textit{Parker Solar Probe} (\textit{PSP}) to exhibit abundant wave activities. The cyclotron wave modes in the sense of ions or electrons are among the most crucial wave components. However, their origin and evolution in the inner heliosphere close to the Sun remain mysteries. Specifically, it remains unknown whether it is an emitted signal from the solar atmosphere or an eigenmode growing locally in the heliosphere due to plasma instability. To address and resolve this controversy, we must investigate the key quantity of the energy change rate of the wave mode. We develop a new technique to measure the energy change rate of plasma waves, and apply this technique to the wave electromagnetic fields measured by \textit{PSP}. We provide the wave Poynting flux in the solar wind frame, identify the wave nature to be the outward propagating fast-magnetosonic/whistler wave mode instead of the sunward propagating waves. We provide the first evidence for growth of the fast-magnetosonic/whistler wave mode in the inner heliosphere based on the derived spectra of the real and imaginary parts of the wave frequencies. The energy change rate rises and stays at a positive level in the same wavenumber range as the bumps of the electromagnetic field power spectral densities, clearly manifesting that the observed fast-magnetosonic/whistler waves are locally growing to a large amplitude.

\end{abstract}



\section{Introduction} \label{sec:intro}

Waves are essential channels of energy conversion in various plasma systems. Particularly for the waves at kinetic scales, wave-particle interaction plays a crucial role in modulating the particles’ velocity distribution, leading to the energization/cooling of plasmas, as well as the kinetic energy transfer between parallel and perpendicular degrees of freedom \citep{marsch2006kinetic,hellinger2006solar, he2015evidence, ruan2016kinetic, howes2017diagnosing, yoon2017kinetic, klein2018majority, verscharen2019multi, duan2020magnetic, verniero2020parker, zhao2020observational}. Regarding space plasmas in the heliosphere, the situation is more complicated. There exist various wave modes: electromagnetic wave modes (e.g., Alfven-cyclotron waves, whistler waves) \citep{jian2009ion, he2011possible, boardsen2015messenger, narita2018space, zhao2018modulation, woodham2019parallel, bowen2020ion, shi2021parker, jagarlamudi2021whistler, zhao2021turbulence}, electrostatic wave modes (e.g., ion-acoustic waves, Langmuir waves) \citep{zhu2019composition, mozer2020large}, and hybrid wave modes (e.g., quasi-perpendicular kinetic Alfv\'en waves) \citep{bale2005measurement, sahraoui2009evidence, he2012reproduction, salem2012identification, chen2013nature, huang2020kinetic}. Observations reveal propagation directions to be anti-sunward or sunward, quasi-parallel or quasi-perpendicular with respect to the local background magnetic field direction. The polarization of the fluctuating vectors (e.g., $\delta \mathbf{B}$, $\delta \mathbf{E}$, $\delta \mathbf{V}$ for the disturbed magnetic, electric, and velocity field vectors, respectively) can be quasi-linear or quasi-circular with left- or right-handedness. It is also desired to distinguish whether the observed waves are dissipative damping or stimulative growing. Therefore, a thorough diagnosis of the kinetic waves in space plasmas, including the solar wind, is undoubtedly a challenging task.

The fluctuating magnetic field can be helpful in determining the propagation but lead to a 180-degree ambiguity. Since the magnetic field is a solenoidal vector field, the wave magnetic field ($\delta\mathbf{B}$) cannot have a component oscillating along the wave vector direction. This feature of the oscillation direction provides a basis for diagnosing the propagation direction. Therefore, approximating the wave vector direction with the minimum variance direction has become one of the main principles when developing wave vector diagnosis methods, such as the MVA method based on time series \citep{sonnerup1967magnetopause}, or the SVD method based on the spectrum or dynamic spectrum \citep{santolik2003singular}. According to these methods, we can preliminarily diagnose whether the wave encountered by a spacecraft has a quasi-parallel propagation or a quasi-perpendicular propagation. For example, we often see that with decreasing wavelength the magnetic compressibility becomes more significant, and the corresponding $\theta_{\rm{kB_0}}$ becomes larger \citep{he2015sunward}. One of the reasons for this change in behavior is the transformation from magnetohydrodynamic (MHD) Alfv\'en waves to kinetic Alfv\'en waves with decreasing scales. However, single-satellite magnetic field measurements can not solve the problem of the 180-degree ambiguity of propagation angle. So, these measurements are unable to judge the real propagation direction of the wave, hence unable to accurately diagnose the nature of wave mode. To unambiguously identify the wave propagation direction, there are two possible solutions: (1) the time delay analysis based on multi-satellite constellation measurements \citep{gershman2017wave}; (2) the consideration of more physical measurements (such as wave electric field, e.g., measured from MMS) \citep{he2019direct, he2020spectra}.

The fluctuating electric field is another crucial variable for wave diagnosis \citep{mozer2013parallel, he2020spectra}. Only when the wave electric and magnetic fields are measured simultaneously, can the wave electromagnetic energy-flux density, that is, the Poynting flux density, be calculated. However, the measurement and calibration of the electric field are more complicated than of the magnetic field due to Debye shielding and the photoelectric effect, which bring a significant challenge to the accurate measurement of the electric field. Fortunately, the number density of the solar wind measured by \textit{PSP} is two orders of magnitude higher than that of the near-Earth solar wind, and the Debye sphere is thus one order of magnitude smaller, making shorter electric-field antennas feasible \citep{bale2016fields, mozer2020dc}. Furthermore, the \textit{PSP} antenna's geometric configuration leaves the potential measurement at the four ends ($U_1$, $U_2$, $U_3$, and $U_4$) unaffected by the wake of the spacecraft. In this way, in the absence of physical adverse factors, the main task for the data analysis is the careful calibration of the electric field. The convection electric field at MHD scales can be used as the benchmark electric field to calibrate the electric field based on multi-point potential measurements \citep{mozer2020dc}. Based on the magnetic field's frozen-in condition at MHD scales, the convection electric field can be approximated by the opposite of the cross product of the velocity and magnetic field vectors ($\mathbf{E}\sim -\mathbf{V}\times\mathbf{B}$). Therefore, the calibration coefficients obtained at MHD scales can be extended to obtain the electric field at kinetic scales. Based on the time series of electric and magnetic fields, it is found that the magnitude of the Poynting vector in the switchback structure is larger than that outside\citep{mozer2020dc}. The reason is that the outflow velocity inside the structure is larger, and so is the angle between the outflow velocity and the magnetic field. Besides, the propagation speed of the kinetic wave's Poynting vector in the heliographic inertial (HGI) reference frame is larger than the solar wind flow speed, suggesting that the wave events under study propagate away from the sun \citep{bowen2020electromagnetic}.

The origin of kinetic scale fluctuations in the solar wind is a controversial topic of research. There are two different views on this issue. (1) One view is that the wave fluctuations are emitted from the solar atmosphere and cascade from the MHD scales to the kinetic scales during their journey of outward propagation \citep{he2009excitation, cranmer2015role, yang2017formation, chandran2019reflection, He2021origin}. (2) The other view is that the kinetic-scale waves are produced locally in the interplanetary space due to some plasma instability \citep{jian2014electromagnetic, wicks2016proton, jiansen2018plasma, zhao2019generation, verniero2020parker}. Because the cascade of the Alfv\'en turbulence preferentially creates anisotropy with $k_{\perp}\gg k_{\parallel}$, the quasi-perpendicular propagation of kinetic Alfv\'en wave may be generated by a cascade along with the outward propagation of MHD waves. The origin mechanism is especially unclear for quasi-parallel kinetic waves (such as ion cyclotron wave or whistler wave). However, due to the frequent existence of spectral peaks, it is generally speculated that these waves are related to the excitation by local instability. In addition, the thermal anisotropy of protons, the beam structures in protons and other ions, and the heat flux caused by the strahl component of the solar wind electrons may cause instability in various plasma states. However, previous studies, which are mainly based on the prediction from linear theory, have not provided direct evidence for the time-varying growth of solar wind kinetic waves.

Therefore, it is one of the cutting-edge frontiers to study and provide evidence of the time-varying evolution (growth or dissipation) of wave events. Quasi-parallel kinetic waves (such as ion cyclotron waves) were once considered an important energy source for solar wind heating. The dissipation of quasi-perpendicular kinetic Alfv\'en waves is also an effective way to heat the solar wind. These viewpoints need to be proved by the direct observation of the dissipation rate spectrum, but the dissipation rate spectrum has been unexplored for a long time. Recently, based on the detection of electromagnetic field and plasma in the magnetosheath turbulence by MMS, the measurement method of the dissipation rate spectrum was proposed \citep{he2019direct}. The dissipation rate spectra of ion cyclotron waves (mainly in the perpendicular direction) and kinetic Alfv\'en waves (mainly in the parallel direction) in magnetosheath turbulence are measured \citep{he2019direct, he2020spectra}. However, the growth rate spectrum of an excited instability has yet to be reported. Although the trivial energy transfer rate from fields to particles as compared with the energy flux density supports local generation scenario of cyclotron waves \citep{vech2020wave}, the direct measurement of the wave growth in the inner heliosphere and the details of the associated growth rate spectrum are still unresolved.

\section{Method of wave diagnosis}
\subsection{Calibration of electric field}
Since the magnetic frozen-in condition holds at MHD scales, the electric field due to convection ($\mathbf{E}=-\mathbf{V}\times\mathbf{B}$) can be viewed and used as the benchmark electric field for the calibration. We regard the calibration procedure for the electric field vector ($E_T$, $E_N$) from the electric potentials measured at four points ($U_1$, $U_2$, $U_3$, $U_4$) as a type of fitting procedure. The input conditions are known as $U_{12}$=$U_2$-$U_1$ and $U_{34}$=$U_4$-$U_3$, and the output variables are $E_T=-(\mathbf{V}\times \mathbf{B})_T$ and $E_N=-(\mathbf{V}\times \mathbf{B})_N$. The fitting parameters to be determined consist of the following parameters: (1) the residual electric potential between the measurement points 1 and 2, (2) the residual electric potential between the measurement points 3 and 4, (3) effective length of the antennas ($L$), (4) the rotation angle $\theta$ from the coordinates defined by the two measurement antennas ($\mathbf{e}_{12}$, $\mathbf{e}_{34}$) to the coordinates (T, N). A set of fitting equations can be obtained based on the known observables and unknown fitting parameters, and written as:
\begin{equation}
    \left[\begin{array}{cc}\cos \theta & \sin \theta \\ \sin \theta & -\cos \theta\end{array}\right]\left[\begin{array}{c}\frac{U_{12}+\Delta U_{12}}{L} \\ \frac{U_{34}+\Delta U_{34}}{L}\end{array}\right]=-\left[\begin{array}{l}(\mathbf{V} \times \mathbf{B})_{T} \\ (\mathbf{V} \times \mathbf{B})_{N}\end{array}\right]
\,\,\, . \label{Eq-1}\end{equation}

To employ the technique of a generalized gradient descent algorithm (``GGDA'') \citep{zhang20122009}, we combine the fitting parameters, and rewrite Equation~\ref{Eq-1} as:
\begin{equation}
    \left[\begin{array}{cccc}U_{12} & U_{34} & 1 / 2 & 1 / 2 \\ \cdots & \cdots & \cdots & \cdots \\ -U_{34} & U_{12} & 1 / 2 & -1 / 2 \\ \cdots & \cdots & \cdots & \cdots\end{array}\right]\left[\begin{array}{c}\frac{\cos \theta}{L} \\ \frac{\sin \theta}{L} \\ C_{1}+C_{2} \\ C_{1}-C_{2}\end{array}\right]=-\left[\begin{array}{c}(\mathbf{V} \times \mathbf{B})_{T} \\ \cdots \\ (\mathbf{V} \times \mathbf{B})_{N} \\ \ldots\end{array}\right]
\,\,\, , \label{Eq-2}\end{equation}
where the fitting parameters are considered as the vector on the left side of Equation~\ref{Eq-2}. The pair of parameters ($C_1$, $C_2$) is expressed as:
\begin{eqnarray}
    C_{1} &=\frac{\cos \theta}{L} \Delta U_{12}+\frac{\sin \theta}{L} \Delta U_{34} \\ 
    C_{2} &=\frac{\sin \theta}{L} \Delta U_{12}-\frac{\cos \theta}{L} \Delta U_{34}
\,\,\,.\label{C1_C2}\end{eqnarray}
If there are $N$ data points in the time sequence, then the sizes of the matrix and the vectors in Equation~\ref{Eq-2} are (4, 2N), (1, 4), (1, 2N). In practice, similar to the time length adopted in \citep{mozer2020dc}, we choose a time window of 12~s as the the time length to employ the fitting approach.      

As the last step of the electric field calibration, we use the fitting parameters derived from the ``GGDA'' to calculate the electric field vectors based on the four-point measurements of the electric potentials at a higher time cadence of 0.0068~second. The calibrated electric field vectors are in the heliographic inertial (HGI) reference frame instead of in the solar wind frame. Moreover, the calibration of $E_{\rm{R}}$ is more complicated than that of $E_{\rm{T}}$ and $E_{\rm{N}}$ since the measurement of the potential $U_5$ is in the wake of the \textit{PSP} spacecraft. Therefore, this work uses mostly $E_{\rm{T}}$ and $E_{\rm{N}}$, and focuses on the parallel/anti-parallel propagating wave events when the local background magnetic field is in the quasi-radial or anti-quasi-radial directions.

\subsection{Formulas of dynamic spectra for Poynting vector, magnetic helicity, and electric field polarization}
We adopt a method similar to that in \citep{podesta2009dependence} to calculate the local background magnetic field ($\mathbf{B}_{\rm{0,local-BG}}$) and the local background flow velocity ($\mathbf{V}_{\rm{sw,local-BG}}$), which are obtained through the convolution between Gaussian windows of different widths and the time sequences of the magnetic vectors and flow velocity vectors. The dynamic spectrum of the Poynting vector in the solar wind frame and its component in the R direction can be calculated as:
\begin{equation}
    \mathbf{PF}(t,p)=\frac{\operatorname{Re}\left(\delta \widetilde{\mathbf{E}^{\prime}} \times \delta \widetilde{\mathbf{B}}^{*}\right)}{\mu_{0}}
\label{PF_tp}\end{equation}
and 
\begin{equation}
    \mathrm{PF}_{R}(t, p)=\frac{Re\left(\delta \widetilde{E^{\prime}}_{\rm{T}} \delta \widetilde{B}_{\rm{N}}^{*}-\delta \widetilde{E^{\prime}}_{\rm{N}} \delta \widetilde{B}_{\rm{T}}^{*}\right)}{\mu_{0}}
\,\,\, , \label{PF_R}\end{equation}
where the independent variables (t, p) represent the time and period, respectively. The complex variables ($\delta \mathbf{E}'$ and $\delta \mathbf{B}$) are the wavelet spectra of electric field (in the reference frame of local background flow) and magnetic field, respectively. The relation between the electric field spectra in the reference frame of local background flow and its counterpart in the heliographic inertial (HGI) reference frame can be expressed as:
\begin{equation}
    \delta \widetilde{\mathbf{E}^{\prime}}=\delta \widetilde{\mathbf{E}}+\mathbf{V}_{\mathrm{sw}, \text { local }-\mathrm{BG}} \times \delta \widetilde{\mathbf{B}}
\,\,\, . \label{E_prime}\end{equation}
Note that, the zero-frequency part of convection electric field as contributed from the convection of the mean magnetic flux by the mean flow ($\mathbf{E}_0=-\mathbf{V}_0\times\mathbf{B}_0$) does not appear in the frequency-dependent Equation~\ref{E_prime}. The normalized and reduced magnetic helicity is calculated according to
\begin{equation}
    \sigma_{\mathrm{m}}(t, p)=+\frac{2 \operatorname{Im}\left(\delta \widetilde{B}_{T} \delta \widetilde{B}_{N}^{*}\right)}{\left|\delta \widetilde{B}_{T}\right|^{2}+\left|\delta \widetilde{B}_{N}\right|^{2}}
\,\,\, , \label{sigma_m}\end{equation}
where $\delta \widetilde{B}_{\rm{T}}$ and $\delta \widetilde{B}_{\rm{N}}$ are the wavelet spectra of the magnetic field components $B_T$ and $B_N$.
Similarly, the “polarization” of the electric field about the R direction can be formulated as:
\begin{equation}
    \sigma_{\mathrm{E}^{\prime}}(t, p)=+\frac{2 \operatorname{Im}\left(\delta \widetilde{E^{\prime}}_{\rm{T}} \delta\widetilde{E^{\prime}}_{\rm{N}}^{*}\right)}{\left|\delta \widetilde{E^{\prime}}_{\rm{T}}\right|^{2}+\left|\delta \widetilde{E^{\prime}}_{\rm{N}}\right|^{2}}
\,\,\, , \label{sigma_E_prime}\end{equation}
where $\delta \widetilde{E}_{\rm{T}}$ and $\delta \widetilde{E}_{\rm{N}}$ represent the wavelet spectra of electric field components in the T and N directions, respectively.

\subsection{Method of identification and classification of wave events}
To identify some ideal events of kinetic waves for further detailed analysis, we propose a set of criteria and list them in Table~1. The variables $\rm{PF}$, $\theta_{\rm{RB}}$, $\sigma_{\rm{m}}$, and $\sigma_{\rm{E'}}$ represent: (1) the Poynting flux density, (2) the angle between radial and local mean magnetic field directions, (3) the normalized reduced magnetic helicity, (4) the polarization wave electric field about the radial direction in the local mean flow frame, respectively. To make sure that the identified wave events possess the typical characteristics of kinetic wave modes, we conduct the following procedure: (1) We select a time window of 30~s to calculate an average of dynamic spectra of the variables ($\rm{PF}$, $\theta_{\rm{RB}}$, $\sigma_{\rm{m}}$, $\sigma_{\rm{E'}}$) at the time scale of 0.3~s. (2) We set the thresholds for the key variables: $\theta^*_{\rm{RB}}=30^\circ$, $\left|\sigma^*_m\right|=0.5$, $\left|\sigma^*_{\rm{E'}}\right|=0.5$.

\begin{table}[h]
\caption{Key variables and their corresponding criteria used for classification of different wave modes (i.e., fast-magnetosonic/whistler mode or Alfv\'en-cyclotron mode propagating sunward or anti-sunward).}
\begin{center}
\begin{tabular}{|l|l|l|l|l|}
\hline
& Poynting Flux & $\theta_{\rm{RB}}$ & $\sigma_{\rm{m}}$  & $\sigma_{\rm{E'}}$                   \\ \hline
\multirow{2}{*}{Anti-sunward FWM} & \multirow{2}{*}{\textgreater{}0} & \textless{}$\theta^*_{\rm{RB}}$ & \textgreater{}+$\left|\sigma^*_{\rm{m}}\right|$ & \textgreater{}+$\left|\sigma^*_{\rm{E'}}\right|$ \\ \cline{3-5} 
&  & \textgreater{}180-$\theta^*_{\rm{RB}}$ & \textless{}-$\left|\sigma^*_{\rm{m}}\right|$  & \textless{}-$\left|\sigma^*_{\rm{E'}}\right|$ \\ \hline
\multirow{2}{*}{Anti-sunward ACM} & \multirow{2}{*}{\textgreater{}0} & \textless{}$\theta^*_{\rm{RB}}$ & \textless{}-$\left|\sigma^*_{\rm{m}}\right|$    & \textless{}-$\left|\sigma^*_{\rm{E'}}\right|$   \\ \cline{3-5} 
& & \textgreater{}180-$\theta^*_{\rm{RB}}$ & \textgreater{}+$\left|\sigma^*_{\rm{m}}\right|$ & \textgreater{}+$\left|\sigma^*_{\rm{E'}}\right|$ \\ \hline
\multirow{2}{*}{Sunward  FWM}     & \multirow{2}{*}{\textless{}0}    & \textless{}$\theta^*_{\rm{RB}}$        & \textless{}-$\left|\sigma^*_{\rm{m}}\right|$    & \textless{}-$\left|\sigma^*_{\rm{E'}}\right|$    \\ \cline{3-5} 
&  & \textgreater{}180-$\theta^*_{\rm{RB}}$ & \textgreater{}+$\left|\sigma^*_{\rm{m}}\right|$ & \textgreater{}+$\left|\sigma^*_{\rm{E'}}\right|$ \\ \hline
\multirow{2}{*}{Sunward ACM}      & \multirow{2}{*}{\textless{}0}    & \textless{}$\theta^*_{\rm{RB}}$        & \textgreater{}+$\left|\sigma^*_{\rm{m}}\right|$ & \textgreater{}+$\left|\sigma^*_{\rm{E'}}\right|$ \\ \cline{3-5} 
&   & \textgreater{}180-$\theta^*_{\rm{RB}}$ & \textless{}-$\left|\sigma^*_{\rm{m}}\right|$    & \textless{}-$\left|\sigma^*_{\rm{E'}}\right|$   \\ \hline
\end{tabular}
\end{center}
\end{table}

\subsection{Estimating the real and imaginary frequencies of wave activity}
Based on a Fourier transform of the Faraday equation, we obtain:
\begin{equation}
    \frac{\omega_{r}+i \gamma}{k} \delta \widetilde{\mathbf{B}}=\hat{\mathbf{e}}_{k} \times \delta \widetilde{\mathbf{E}^{\prime}}
\,\,\, .\label{FourierTransform_of_FaradayEq}
\end{equation}
If the wave is a transverse wave with both electric and magnetic field fluctuations oscillating in the directions perpendicular to the wave vector, as it is the case in quasi-parallel propagating Alfv\'en/ion-cyclotron waves and fast-magnetosonic/whistler waves for examples, Equation~\ref{FourierTransform_of_FaradayEq} can be rewritten as
\begin{equation}
    \frac{\omega_{r}+i \gamma}{k}=\frac{\delta \widetilde{\mathbf{E}^{\prime}} \times \delta \widetilde{\mathbf{B}}^{*}}{\delta \widetilde{\mathbf{B}} \cdot \delta\widetilde{\mathbf{B}}^{*}}
\,\,\, .\label{omega_plus_i_gamma_over_k}\end{equation}

Therefore, based on the wavelet spectra of the electric and magnetic field, we obtain the dynamic spectrum of dispersion relation and growth rate
\begin{equation}
    \left(\begin{array}{c}\frac{\omega}{k} \\ \frac{\gamma}{k}\end{array}\right)=\left(\begin{array}{c}\operatorname{Re}\left(\frac{\delta \widetilde{\mathbf{E}^{\prime}} \times \delta \widetilde{\mathbf{B}}^{*}}{\delta \widetilde{\mathbf{B}} \cdot \delta \widetilde{\mathbf{B}}^{*}}\right) \\ \operatorname{Im}\left(\frac{\delta \widetilde{\mathbf{E}^{\prime}} \times \delta \widetilde{\mathbf{B}}^{*}}{\delta \widetilde{\mathbf{B}} \cdot \delta \widetilde{\mathbf{B}}^{*}}\right)\end{array}\right)
\,\,\, .\label{omega_k_gamma_k}\end{equation}
We note that, the above equation is a simplified version for the situation of quasi-parallel transverse waves. In general, the wave group speed is determined by the ratio of energy flux density to the energy density, with the energy flux density being the sum of Poynting flux and kinetic flux and the energy density being contributed by the fluctuating electromagnetic field energy and plasma kinetic energy \citep{stix1992waves, swanson2003plasma}. According to the Doppler-shift effect caused by the solar wind flow, the relation between wave frequencies ($\omega_{\rm{sc}}$) in the spacecraft reference frame and in the solar wind flow reference frame ($\omega_{\rm{pl}}$) can be expressed as:
\begin{equation}
    \frac{\omega_{\mathrm{sc}}}{k}=\frac{\omega_{\mathrm{pl}}}{k}+V_{\mathrm{sw}} \cos \theta_{k V}
\,\,\, ,\label{omega_sc_k}\end{equation}
where $V_{\rm{sw}}$ is the local background solar wind flow velocity, and $\theta_{\rm{kV}}$ is the angle between $V_{\rm{sw}}$ and the wave vector $\mathbf{k}$. The direction of the wave vector can be determined without the problem of 180-degree ambiguity by considering the analysis result from the ``singular value decomposition'' (SVD) method and the direction of the Poynting vector relative to the background magnetic field. For convenience, hereafter, we drop the subscript ``pl'' in ``$\omega_{\rm{pl}}$'' for simplicity. Based on Equations~(\ref{omega_k_gamma_k}) and~(\ref{omega_sc_k}), we further derive formulas of $k$, $\omega$, and $\gamma$, which read as
\begin{equation}
    k=\omega_{\mathrm{sc}} \frac{1}{\omega / k+V_{\mathrm{sw}} \cos \theta_{k V}}
\,\,\, , \label{k}\end{equation}
\begin{equation}
    \omega=\omega_{\mathrm{sc}} \frac{\omega / k}{\omega / k+V_{\mathrm{sw}} \cos \theta_{k V}}
\,\,\, ,\label{omega}\end{equation}
\begin{equation}
    \gamma=\omega_{\mathrm{sc}} \frac{\gamma / k}{\omega / k+V_{\mathrm{sw}} \cos \theta_{k V}}
\,\,\, .\label{gamma}\end{equation}


To validate the credibility of applying Equation~(\ref{PF_R}) to the measurement of the Poynting vector, we also propose a formula for calculating the phase difference between the wave electric field $\delta \mathbf{E}'$ ($\phi(\delta \mathbf{E}')$) and the wave magnetic field $\delta \mathbf{B}$ ($\phi(\delta \mathbf{B})$) (see Equation~(\ref{phi_dB_dE})), and calculate its distribution in the time and scale dimensions.
\begin{equation}
    \phi\left(\delta \mathbf{B}_{\perp}, \delta \mathbf{E}_{\perp}^{\prime}\right)=\phi\left(\delta \mathbf{B}_{\perp}\right)-\phi\left(\delta \mathbf{E}_{\perp}^{\prime}\right)
\,\,\, ,\label{phi_dB_dE}\end{equation}
where both $\delta B_{\perp}$ and $\delta E'_{\perp}$ correspond to the wavelet decomposition of the original time sequences at different scales. The polarity of the Poynting vector can be inferred from the phase difference: (1) $\delta \mathbf{E}'\times \delta \mathbf{B}$ is positive for $\phi\left(\delta \mathbf{B}_{\perp}, \delta \mathbf{E}_{\perp}^{\prime}\right) \in(0,180)^{\circ}$ ; (2) $\delta \mathbf{E}'\times \delta \mathbf{B}$ is negative for $\phi\left(\delta \mathbf{B}_{\perp}, \delta \mathbf{E}_{\perp}^{\prime}\right) \in(-180,0)^{\circ}$.

\section{Event Analysis}
\subsection{Analysis Steps}
We conduct the search and analysis of interesting wave events based on the measurements from \textit{PSP} during its first encounter on November 4, 2018. We break this task into six steps. 
(1) The first step is to calibrate the electric field segment by segment according to Equation~(\ref{Eq-2}), and thereby realizing the conversion from the four-point electric potentials to the 2D electric field vectors.
(2) We then invoke Equation~(\ref{E_prime}) to realize the coordinate transformation of the electric field from the spacecraft reference frame to the reference frame of the local solar wind background flow.
(3) We calculate the dynamic spectrum of the Poynting flux along the R-direction according to Equation~(\ref{PF_R}).
(4) We calculate the dynamic spectrum of the magnetic helicity and electric polarization about the R-direction with Equations (\ref{sigma_m}) and (\ref{sigma_E_prime}), respectively.
(5) We calculate $\omega/k$ and $\gamma/k$ for the wave events.
(6) We estimate the wave number, real part and imaginary part of wave frequency according to Equations (\ref{k}-\ref{gamma}).
We classify the wave events based on the analysis results of the above first five steps and as per Table-1. In this way, we accomplish the goal of diagnosing the key characteristics (e.g., propagation direction, the polarization, and the growth/damping rate) of the wave events. 

\subsection{Power spectral densities and polarization of wave electromagnetic fields}
As a typical example, we show a wave event of outward propagation, right-hand polarization about $\mathbf{B}_{\rm{0}}$, and positive growth. The time interval of this event is between [18:28, 18:31]~UT on Nov 4, 2018. In Figure~1a and~1b, we display and compare the calibrated electric field ($E_{\rm{T}}$, $E_{\rm{N}}$) and the induced electric field based on the measurements of magnetic field and bulk velocity ($-(\mathbf{V}\times \mathbf{B})_{\rm{T}}$, $-(\mathbf{V}\times \mathbf{B})_{\rm{N}}$). The two types of electric field match well with one another. Therefore, we use the calibrated electric field to analyze the propagation direction and growth/damping rate of the observed wave. We apply wavelet decomposition to the time sequences of the electric and magnetic field components ($E_{\rm{T}}$, $E_{\rm{N}}$, $B_{\rm{T}}$, $B_{\rm{N}}$), and obtain the corresponding band-pass waves in the frequency range of [0.2, 10]~Hz, which are illustrated in Figure 1c-1f, respectively. To further diagnose how the wave propagates in the solar wind reference frame, we transform the electric field from the spacecraft reference frame to the solar wind reference frame.  

\begin{figure}
\plotone{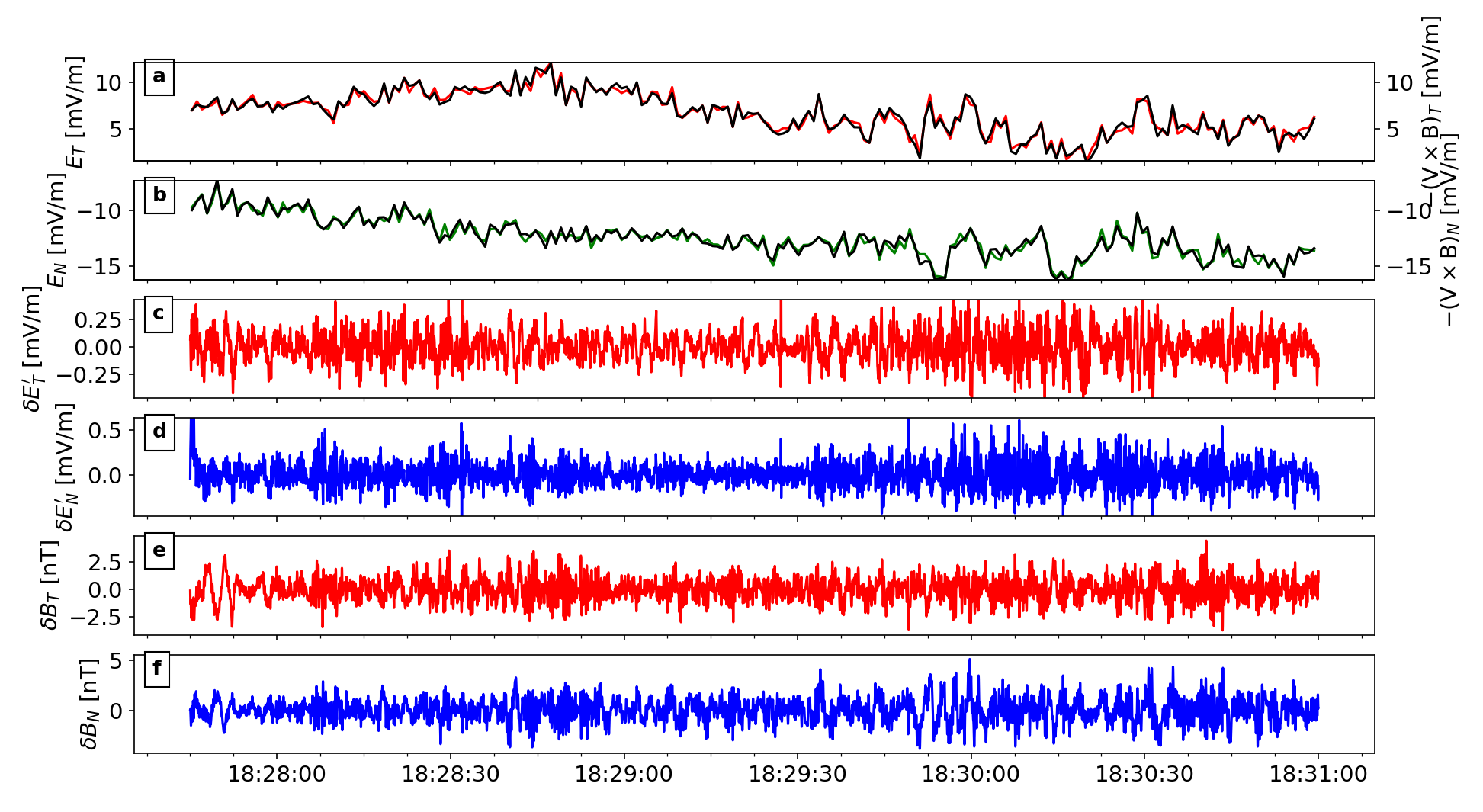}
\caption{Time sequences of the electric and magnetic field for a wave event in the solar wind measured by \textit{PSP} during its first encounter. (a) Consistency between the calibrated $E_{\rm{T}}$ from four-point electric potential differences (red) and the calculated $E_{\rm{T}}$ from $-\mathbf{V}\times\mathbf{B}$ (black). (b) Good match between the calibrated $E_{\rm{N}}$ (green) and the calculated $E_{\rm{N}}$ (black). (c \& d) The band-pass ($f_{\rm{SC}}\in[0.2, 10]$~Hz) wave fluctuations of $E_{\rm{T}}$ and $E_{\rm{N}}$ components in the local solar wind background flow frame ($\delta E'_{\rm{T}}$ and $\delta E'_{\rm{N}}$). (e \& f) The band-pass ($f_{\rm{SC}}\in [0.2, 10]$~Hz) wave fluctuations of the $B_{\rm{T}}$ and $B_{\rm{N}}$ components.\label{fig:fig1}}
\end{figure}

We conduct a detailed analysis of the magnetic field (including the local background and the fluctuating magnetic field) and the electric field (the fluctuating electric field in the local solar wind background frame). We find that the local background magnetic field direction is mainly sunward with $\theta_{BR}\gtrsim 140^\circ$ (see Figure~2a). The magnetic field fluctuations are mainly in the transverse directions, indicating the state of approximate incompressibility ($PSD(\delta B_{\perp})$ in Figure~2b is dominant over $PSD(\delta B_{\parallel})$ in Figure~2c). For most times of the interval, there are evident enhanced signals of $PSD(\delta B_{\perp})$ at periods of [0.2, 0.4]~s (see Figure~2b). Since $B_{\rm{0,local}}$ is quasi-anti-parallel to the R direction, the T and N directions can be approximated as the two directions perpendicular to $B_{\rm{0,local}}$, rendering convenience for the analysis of the transverse wave electric field. We clearly identify the enhanced signals of $PSD(\delta E'_t)$ and $PSD(\delta E'_n)$ in the same period range of [0.2, 0.4]~s (Figure 2d \& 2e). The magnetic helicity spectrum as calculated according to Equation~\ref{sigma_m} shows obvious negative polarity in the period range of [0.2, 0.4]~s (see Figure 2f). Likewise, the polarization of the wave electric field (in the local solar wind background frame) around the R direction, which is calculated from Equation~\ref{sigma_E_prime}, appears with negative polarity (Figure~2g). The good match between magnetic helicity and electric polarization indicates the high-quality measurements of the electric and magnetic fields of this wave event, which can be further analyzed to investigate its propagation direction and activity of growth/dissipation.

\begin{figure}
\includegraphics[width=0.95\textwidth]{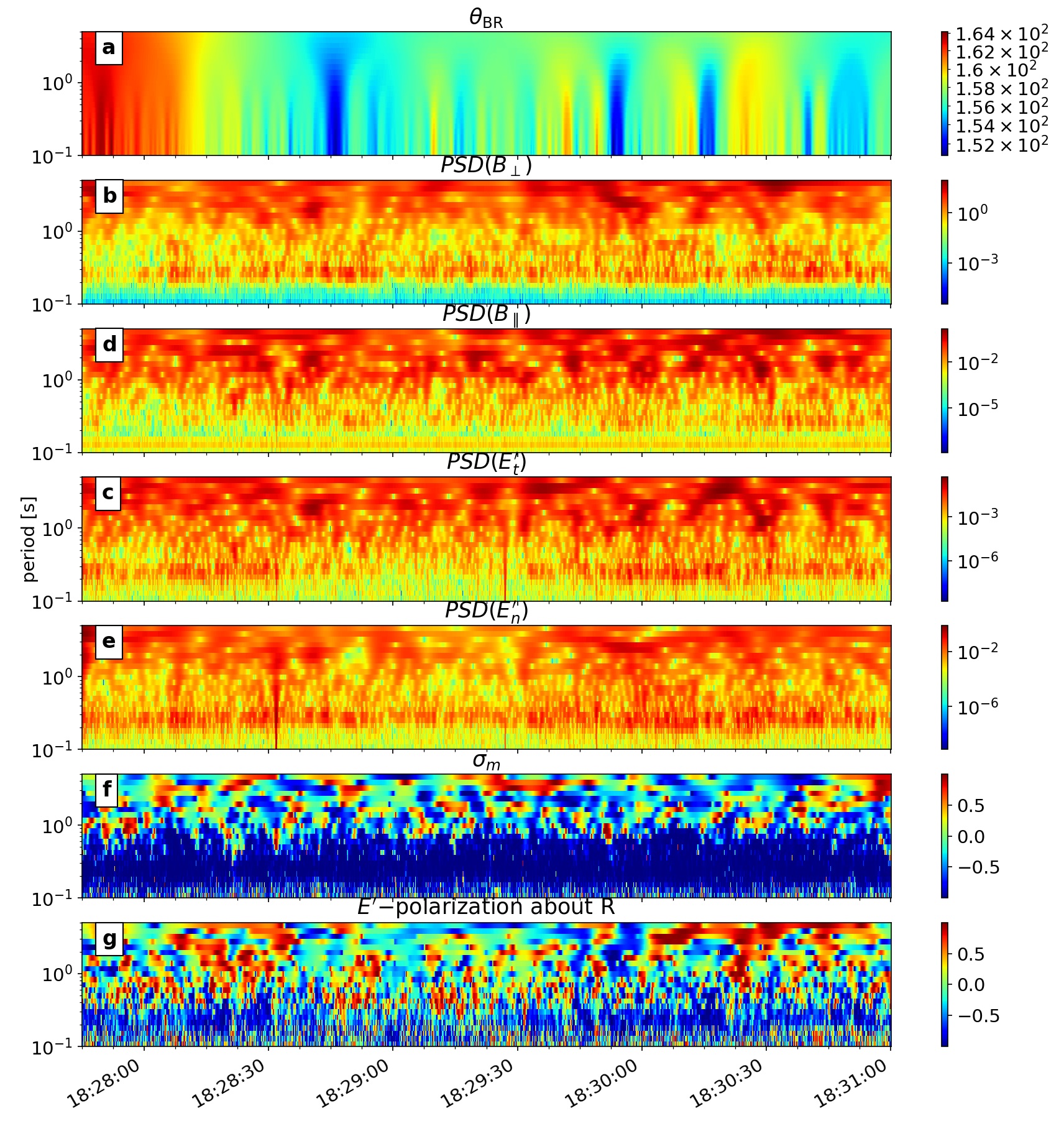}
\caption{Dynamic spectra of the magnetic and electric fields. (a) Time-period distribution of $\theta_{\rm{BR}}$, the angle between the local background magnetic field direction and the radial direction. (b \& c) Time-period distribution of power spectral densities of transverse and longitudinal magnetic field components ($PSD(\delta B_{\perp})$ and $PSD(\delta B_{\parallel})$). (d \& e) Time-period distribution of $PSD(\delta E'_{\rm{T}})$ and $PSD(\delta E'_{\rm{N}})$ in the local solar wind background frame. (f) Time-period distribution of $\sigma_{\rm{m}}$. (g) Time-period distribution of the electric field polarization about the R direction. \label{fig:fig2}}
\end{figure}

\section{Diagnosis of Propagation and Evolution of Wave Events}
According to Equation~\ref{PF_R}, we calculate and illustrate the dynamic spectrum of the Poynting flux density in the R direction ($PF_{\rm{R}}$) (see Figure~3a). During the time interval of [18:27:45, 18:31:00] and in the period range of [0.1, 0.5]~s, $PF_{\rm{R}}$ is basically greater than 0, suggesting that the waves propagate outward quasi-anti-parallel to the sunward local background magnetic field direction. Referring to Table~1, we clearly identify this wave event as outward propagating fast-magnetosonic/whistler waves with right-hand polarization of electromagnetic field vectors about the background magnetic field direction. In Figure~3b, we observe that the phase angle differences, $\phi\left(\delta \mathbf{B}_{\perp}, \delta \mathbf{E}_{\perp}^{\prime}\right) \in(0,180)^{\circ}$ and $\phi\left(\delta \mathbf{B}_{\perp}, \delta \mathbf{E}_{\perp}^{\prime}\right) \in(-180,0)^{\circ}$ correspond to $PF_{\rm{R}}> 0$ and $PF_{\rm{R}}<0$ in Figure~3a, respectively. 

We calculate the dynamic spectra of $\omega/k$ and $\gamma/k$ according to Equation~\ref{omega_k_gamma_k} (see Figure 3c and 3d). Moreover, we calculate the dynamic spectral distribution of $\gamma/\left|\omega\right|$ and $\gamma$ according to Equation~\ref{gamma} (see Figure~3e and 3f). In the case of our study, we approximate $\theta_{kV}$ in Equation~\ref{gamma} with $0^\circ$ since $\theta_{kB}\sim 180^\circ$ and $\theta_{BV}\sim 180^\circ$ during the interval of study. We can see that $\gamma$ is most of the time greater than 0 in the time-period distribution, especially in the period range of [0.1, 0.5]~s. This evidence strongly suggests that the observed fast-magnetosonic/whistler waves are growing during the time of observation. 

\begin{figure}
\includegraphics[width=0.95\textwidth]{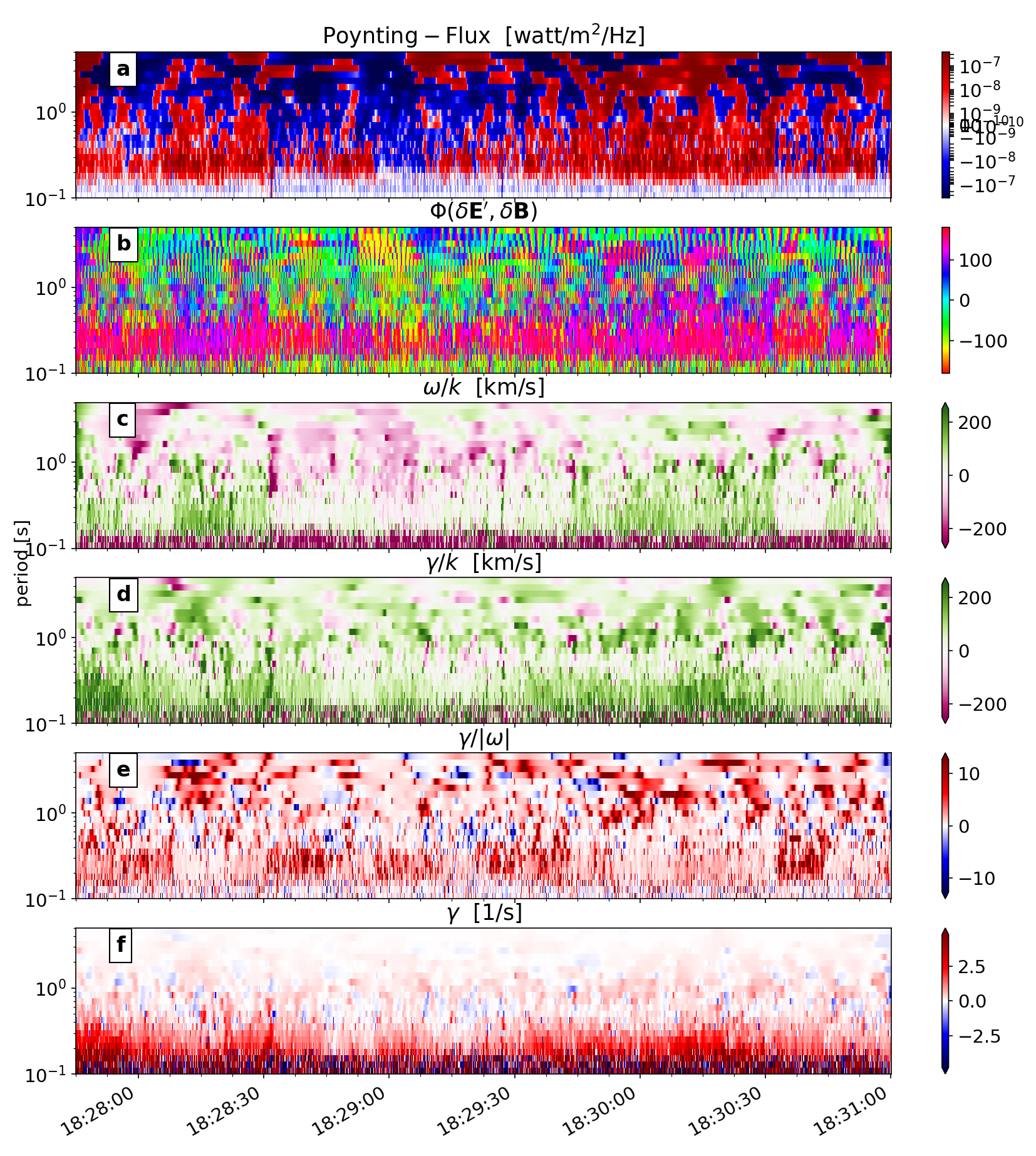}
\caption{Analysis result of wave propagation and growth/damping. (a) Time-period distribution (dynamic spectrum) of the Poynting flux density component $PF_{\rm{R}}$. (b) Time-period distribution of $\phi(\delta E'_\perp, \delta B_\perp)$ (=$\phi(\delta B_\perp)-\phi(\delta E'_\perp)$). (c) Dynamic spectra of $\omega/k$. (d) Dynamic spectra of $\gamma/k$. (e) Dynamic spectra of $\gamma/|\omega|$. (f) Dynamic spectra of $\gamma$. \label{fig:fig3}}
\end{figure}

We apply a further statistical analysis of the above results of wave propagation and growth. We select seven time scales ($\tau$=0.141, 0.167, 0.197, 0.234, 0.277, 0.329, 0.390~s), and count the value-dependent occurrence frequency distribution of multiple variables (e.g., $PF_{\rm{R}}$, $\phi(\delta E'_\perp, \delta B_\perp)$, $\omega/k$, $\gamma/k$, $\gamma/|\omega|$, $\gamma$) (see Figure 4a-4f). At scales shorter than 0.5~s, $PF_{\rm{R}}$ appears more on the positive side, $\phi(\delta E'_\perp, \delta B_\perp)$ appears more in the angle range of (0, 180) degrees. The distribution of $\gamma$ is asymmetric, with more intervals on the side greater than 0, indicating the nature of local excitation and emission for the studied fast-magnetosonic/whistler waves.  

\begin{figure}
\includegraphics[width=0.90\textwidth]{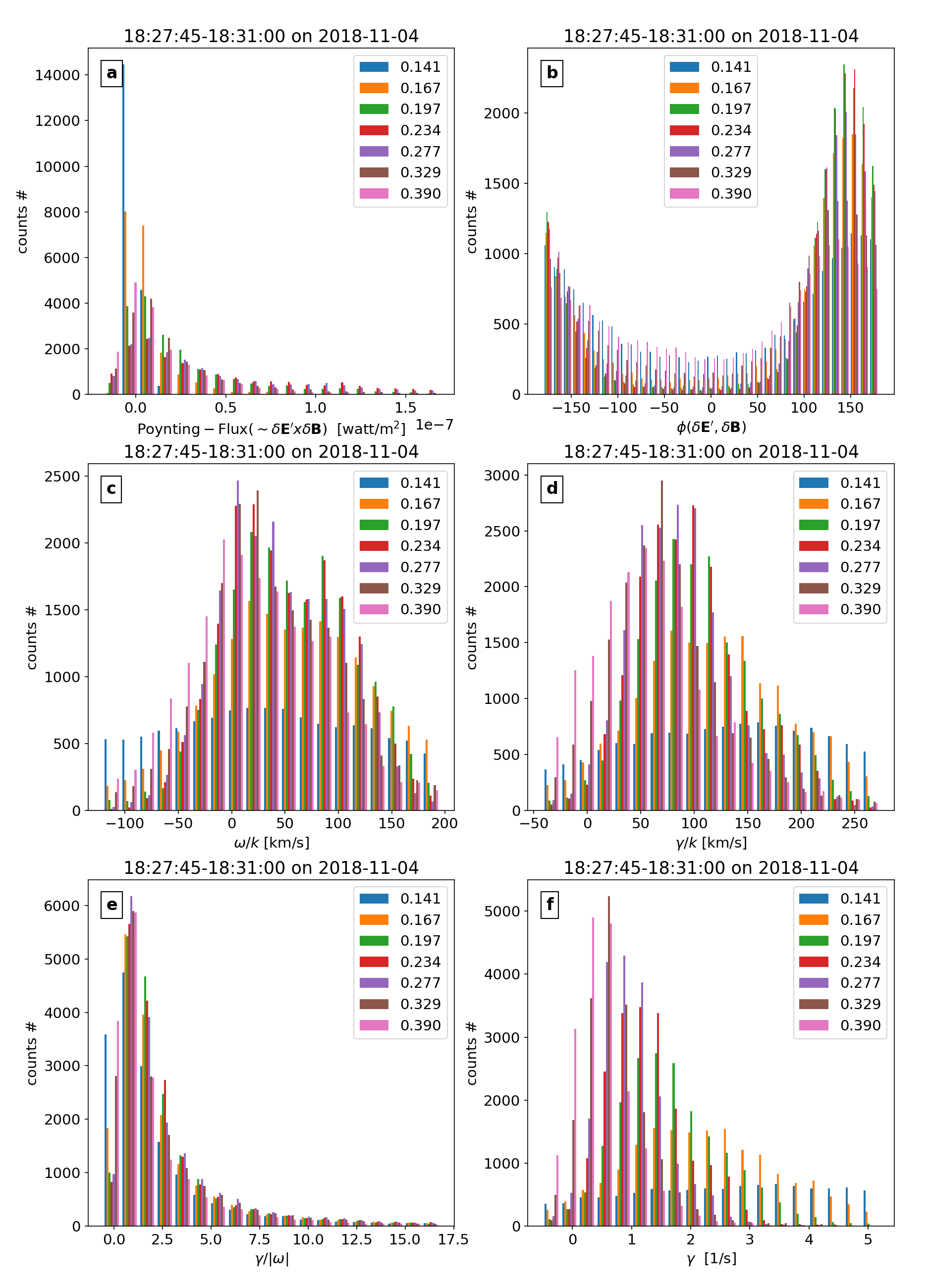}
\caption{Occurrence frequency distribution (i.e., un-normalized probability distribution function) of multiple variables related with wave propagation and evolution at seven different time scales ($\tau$=0.141, 0.167, 0.197, 0.234, 0.277, 0.329, 0.390~s). (a) Frequency histogram of $PF_{\rm{R}}$ component. (b) Frequency histogram of $\phi(\delta E'_\perp, \delta B_\perp)$. (c) Frequency histogram of $\omega/k$. (d) Frequency histogram of $\gamma/k$. (e) Frequency histogram of $\gamma/|\omega|$. (f) Frequency histogram of $\gamma$. \label{fig:fig4}}
\end{figure}

To view the variation of $\rm{PSD}(\delta B_\perp)$, $\rm{PSD}(\delta E'_{\rm{T,N}})$, $\omega$ and $\gamma$ as a function of $f_{\rm{sc}}$ from a statistical perspective, we plot the occurrence frequency distribution in the 2D space of ($f_{\rm{sc}}$, $\rm{PSD}(\delta B_\perp)$), ($f_{\rm{sc}}$, $\rm{PSD}(\delta E'_{\rm{T,N}})$), ($f_{\rm{sc}}$, $\omega$), and ($f_{\rm{sc}}$, $\gamma$) (see Figure 5). We can see from Figure 5a and 5b that, both $\rm{PSD}(\delta B_\perp)$ and $\rm{PSD}(\delta E'_{\rm{T,N}})$ show an obvious spectral bump around $f_{\rm{sc}}\sim 0.4$ Hz. Such spectral bump structure indicates that, the wave signal is stronger than the background turbulence level, probably due to its excitation and unstable growth. Unlike for damped or freely propagating waves, the growth rate ($\gamma$) of the active wave evidently exceeds the zero level (see Figure~5d), and even approaches a level comparable to the derived wave frequency (Figure~5c), offering further direct evidence that the active wave is growing during the time of the observation. At higher frequency beyond the PSD's bump, the occurrence frequency distributions of both $\omega$ and $\gamma$ become diffusive (see the right end of Figures~5c\&5d), probably due to the uncertainty of the electric field measurements at higher frequency.

\begin{figure}
\includegraphics[width=0.95\textwidth]{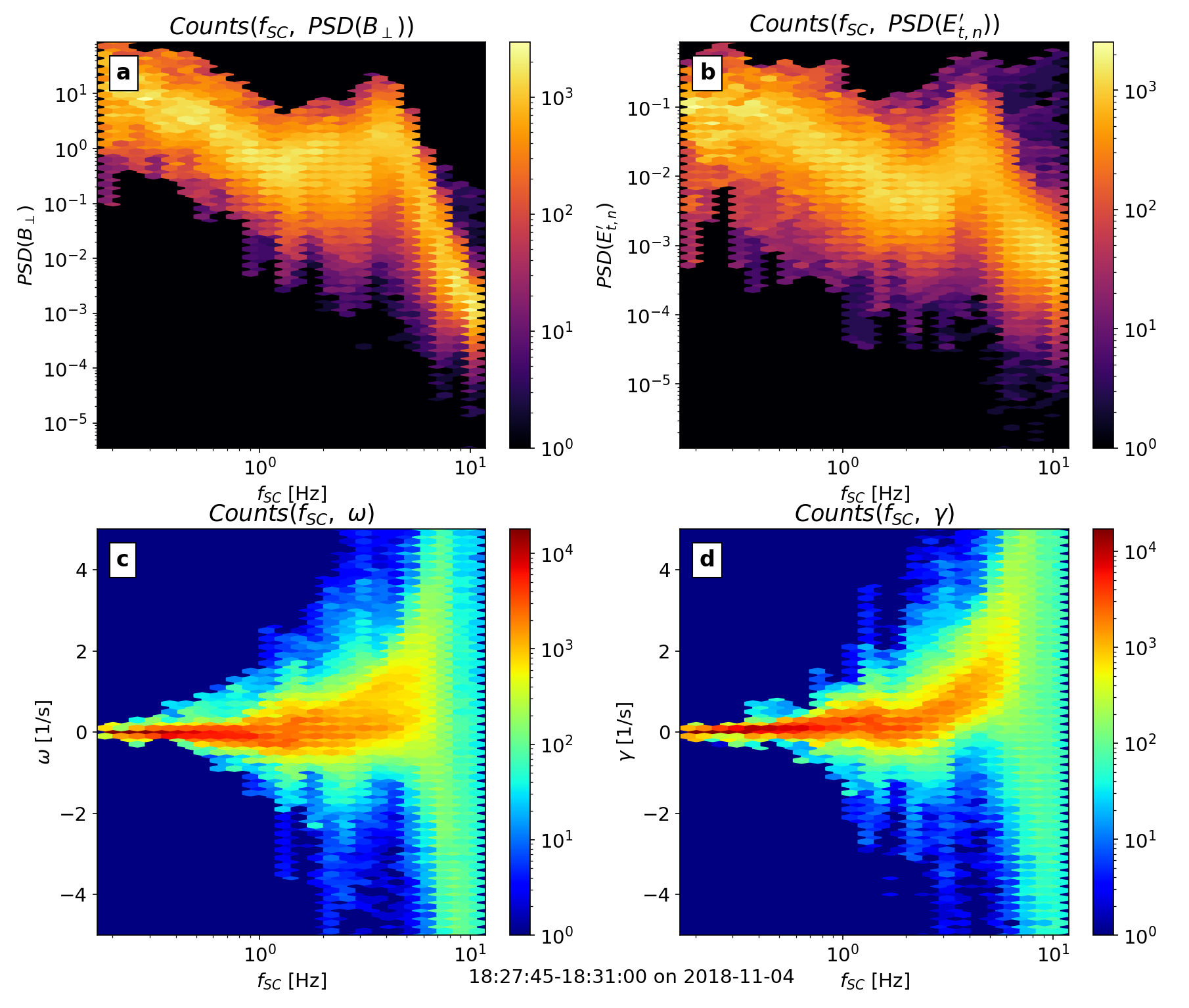}
\caption{Evidence of fast-magnetosonic/whistler wave growth leading to amplitude enhancement. (a \& b) Occurrence frequency distribution of $\rm{PSD}(\delta B_\perp)$ and $\rm{PSD}(\delta E'_{\rm{T,N}})$ at various frequencies in the spacecraft frame. (c \& d) Occurrence frequency distribution of $\omega$ and $\gamma$ at various frequencies in the spacecraft frame.\label{fig:fig5}}
\end{figure}

\section{Discussion and conclusions}
In this work, we propose a method to quantify the energy flux density of wave propagation (i.e., Poynting flux density for electromagnetic waves) and the growth/dissipation rate spectrum. Based on this method, we further put forward a set of diagnosis criteria for the nature of kinetic wave events in the heliosphere. We apply this analysis method and the diagnosis criteria to in situ measurements from \textit{PSP} in the inner heliosphere. As an example, we identify an event of outward propagating fast-magnetosonic/whistler waves with right-hand polarization. For this wave event, we provide the dynamic spectra of physical variables (power spectral densities, magnetic helicity, electric field polarization, Poynting flux density, phase difference between electric and magnetic fields, wave frequency, and normalized rate of change of the wave energy density). We find that the wave event is not in a time-steady state but in the temporally growing phase, evidenced by the positive bump of the $\gamma(f_{\rm{sc}})$ spectral profile, which is physically responsible for the spectral bumps appearing on the PSDs of the electric and magnetic field fluctuations.

This work addresses the issue of the origin of kinetic waves in the inner heliosphere. We point out that kinetic waves are not necessarily created in the solar wind source region, though some proportion of waves may be launched from the solar atmosphere through magnetic reconnection or turbulent advection shaking \citep{He2021origin, zank2020origin}. Instead they can be locally excited and grow due to instability in the inner heliosphere. The results of this work indicate that the inner heliosphere shall be regarded as a critical region for the birth and development of kinetic waves. This suggests that the inner heliosphere exhibits complex wave-particle coupling processes, involving the velocity distributions of various plasma species and the time-varying evolution of different wave modes. The free energy responsible for the fast-magnetosonic/whistler waves may come from the drift ion population, electron heat flux, and electron thermal anisotropy \citep{verscharen2013instabilities, stansby2016experimental, narita2016electron, tong2019statistical, sun2020electron}. In the future, we require a combination of both the electromagnetic field information and the particle phase space density to explore the mystery of kinetic waves and their wave-particle interactions in the inner heliosphere in a comprehensive way.


\begin{acknowledgments}
We thank the NASA Parker Solar Probe Mission and the FIELDS and SWEAP teams for use of data. PSP data is available on SPDF (https://cdaweb.sci.gsfc.nasa.gov/index.html/). The work at Peking University is supported by NSFC under contracts 41674171 and 41874200, and by CNSA under contracts No. D020301 and D020302. D.V. from UCL is supported by the STFC Ernest Rutherford Fellowship 354 ST/P003826/1 and STFC Consolidated Grant ST/S000240/1. G.Q. Zhao is supported by NSFC under contract 41874204 and partly by the Project for Scientific Innovation Talent in Universities of Henan Province (19HASTIT020).
\end{acknowledgments}

\bibliography{ACW_Growth_PSP}{}

\begin{thebibliography}{}
\expandafter\ifx\csname natexlab\endcsname\relax\def\natexlab#1{#1}\fi
\providecommand{\url}[1]{\href{#1}{#1}}
\providecommand{\dodoi}[1]{doi:~\href{http://doi.org/#1}{\nolinkurl{#1}}}
\providecommand{\doeprint}[1]{\href{http://ascl.net/#1}{\nolinkurl{http://ascl.net/#1}}}
\providecommand{\doarXiv}[1]{\href{https://arxiv.org/abs/#1}{\nolinkurl{https://arxiv.org/abs/#1}}}

\bibitem[{Bale {et~al.}(2005)Bale, Kellogg, Mozer, Horbury, \&
  Reme}]{bale2005measurement}
Bale, S., Kellogg, P., Mozer, F., Horbury, T., \& Reme, H. 2005, Physical
  Review Letters, 94, 215002

\bibitem[{Bale {et~al.}(2016)Bale, Goetz, Harvey, Turin, Bonnell, De~Wit,
  Ergun, MacDowall, Pulupa, Andr{\'e}, {et~al.}}]{bale2016fields}
Bale, S., Goetz, K., Harvey, P., {et~al.} 2016, Space science reviews, 204, 49

\bibitem[{Boardsen {et~al.}(2015)Boardsen, Jian, Raines, Gershman, Zurbuchen,
  Roberts, \& Korth}]{boardsen2015messenger}
Boardsen, S., Jian, L., Raines, J., {et~al.} 2015, Journal of Geophysical
  Research: Space Physics, 120, 10

\bibitem[{Bowen {et~al.}(2020{\natexlab{a}})Bowen, Mallet, Huang, Klein,
  Malaspina, Stevens, Bale, Bonnell, Case, Chandran, {et~al.}}]{bowen2020ion}
Bowen, T.~A., Mallet, A., Huang, J., {et~al.} 2020{\natexlab{a}}, The
  Astrophysical Journal Supplement Series, 246, 66

\bibitem[{Bowen {et~al.}(2020{\natexlab{b}})Bowen, Bale, Bonnell, Larson,
  Mallet, McManus, Mozer, Pulupa, Vasko, Verniero,
  {et~al.}}]{bowen2020electromagnetic}
Bowen, T.~A., Bale, S.~D., Bonnell, J., {et~al.} 2020{\natexlab{b}}, The
  Astrophysical Journal, 899, 74

\bibitem[{Chandran \& Perez(2019)}]{chandran2019reflection}
Chandran, B.~D., \& Perez, J.~C. 2019, Journal of Plasma Physics, 85

\bibitem[{Chen {et~al.}(2013)Chen, Boldyrev, Xia, \& Perez}]{chen2013nature}
Chen, C., Boldyrev, S., Xia, Q., \& Perez, J. 2013, Physical review letters,
  110, 225002

\bibitem[{Cranmer {et~al.}(2015)Cranmer, Asgari-Targhi, Miralles, Raymond,
  Strachan, Tian, \& Woolsey}]{cranmer2015role}
Cranmer, S.~R., Asgari-Targhi, M., Miralles, M.~P., {et~al.} 2015,
  Philosophical Transactions of the Royal Society A: Mathematical, Physical and
  Engineering Sciences, 373, 20140148

\bibitem[{Duan {et~al.}(2020)Duan, He, Wu, \& Verscharen}]{duan2020magnetic}
Duan, D., He, J., Wu, H., \& Verscharen, D. 2020, The Astrophysical Journal,
  896, 47

\bibitem[{Gershman {et~al.}(2017)Gershman, Adolfo, Dorelli, Boardsen, Avanov,
  Bellan, Schwartz, Lavraud, Coffey, Chandler, {et~al.}}]{gershman2017wave}
Gershman, D.~J., Adolfo, F., Dorelli, J.~C., {et~al.} 2017, Nature
  communications, 8, 1

\bibitem[{He {et~al.}(2009)He, Marsch, Tu, \& Tian}]{he2009excitation}
He, J., Marsch, E., Tu, C., \& Tian, H. 2009, The Astrophysical Journal
  Letters, 705, L217

\bibitem[{He {et~al.}(2011)He, Marsch, Tu, Yao, \& Tian}]{he2011possible}
He, J., Marsch, E., Tu, C., Yao, S., \& Tian, H. 2011, The Astrophysical
  Journal, 731, 85

\bibitem[{He {et~al.}(2015{\natexlab{a}})He, Pei, Wang, Tu, Marsch, Zhang, \&
  Salem}]{he2015sunward}
He, J., Pei, Z., Wang, L., {et~al.} 2015{\natexlab{a}}, The Astrophysical
  Journal, 805, 176

\bibitem[{He {et~al.}(2012)He, Tu, Marsch, \& Yao}]{he2012reproduction}
He, J., Tu, C., Marsch, E., \& Yao, S. 2012, The Astrophysical Journal, 749, 86

\bibitem[{He {et~al.}(2015{\natexlab{b}})He, Wang, Tu, Marsch, \&
  Zong}]{he2015evidence}
He, J., Wang, L., Tu, C., Marsch, E., \& Zong, Q. 2015{\natexlab{b}}, The
  Astrophysical Journal Letters, 800, L31

\bibitem[{He {et~al.}(2020)He, Zhu, Verscharen, Duan, Zhao, \&
  Wang}]{he2020spectra}
He, J., Zhu, X., Verscharen, D., {et~al.} 2020, The Astrophysical Journal, 898,
  43

\bibitem[{He {et~al.}(2021)He, Zhu, Yang, Hou, Duan, Zhang, \&
  Wang}]{He2021origin}
He, J., Zhu, X., Yang, L., {et~al.} 2021, The Astrophysical Journal Letters,
  913, L14, \dodoi{10.3847/2041-8213/abf83d}

\bibitem[{He {et~al.}(2019)He, Duan, Wang, Zhu, Li, Verscharen, Wang, Tu,
  Khotyaintsev, Le, {et~al.}}]{he2019direct}
He, J., Duan, D., Wang, T., {et~al.} 2019, The Astrophysical Journal, 880, 121

\bibitem[{Hellinger {et~al.}(2006)Hellinger, Tr{\'a}vn{\'\i}{\v{c}}ek, Kasper,
  \& Lazarus}]{hellinger2006solar}
Hellinger, P., Tr{\'a}vn{\'\i}{\v{c}}ek, P., Kasper, J.~C., \& Lazarus, A.~J.
  2006, Geophysical research letters, 33

\bibitem[{Howes {et~al.}(2017)Howes, Klein, \& Li}]{howes2017diagnosing}
Howes, G.~G., Klein, K.~G., \& Li, T.~C. 2017, Journal of Plasma Physics, 83

\bibitem[{Huang {et~al.}(2020)Huang, Zhang, Sahraoui, He, Yuan, Andr{\'e}s,
  Hadid, Deng, Jiang, Yu, {et~al.}}]{huang2020kinetic}
Huang, S., Zhang, J., Sahraoui, F., {et~al.} 2020, The Astrophysical Journal
  Letters, 897, L3

\bibitem[{Jagarlamudi {et~al.}(2021)Jagarlamudi, de~Wit, Froment,
  Krasnoselskikh, Larosa, Bercic, Agapitov, Halekas, Kretzschmar, Malaspina,
  {et~al.}}]{jagarlamudi2021whistler}
Jagarlamudi, V., de~Wit, T.~D., Froment, C., {et~al.} 2021, Astronomy \&
  Astrophysics

\bibitem[{Jian {et~al.}(2014)Jian, Wei, Russell, Luhmann, Klecker, Omidi,
  Isenberg, Goldstein, Figueroa-Vi{\~n}as, \&
  Blanco-Cano}]{jian2014electromagnetic}
Jian, L., Wei, H., Russell, C., {et~al.} 2014, The Astrophysical Journal, 786,
  123

\bibitem[{Jian {et~al.}(2009)Jian, Russell, Luhmann, Strangeway, Leisner, \&
  Galvin}]{jian2009ion}
Jian, L.~K., Russell, C.~T., Luhmann, J.~G., {et~al.} 2009, The Astrophysical
  Journal Letters, 701, L105

\bibitem[{Jiansen {et~al.}(2018)Jiansen, Xingyu, Yajie, Chadi, Michael, Hui,
  Wenzhi, Lei, \& Chuanyi}]{jiansen2018plasma}
Jiansen, H., Xingyu, Z., Yajie, C., {et~al.} 2018, The Astrophysical Journal,
  856, 148

\bibitem[{Klein {et~al.}(2018)Klein, Alterman, Stevens, Vech, \&
  Kasper}]{klein2018majority}
Klein, K., Alterman, B., Stevens, M., Vech, D., \& Kasper, J. 2018, Physical
  review letters, 120, 205102

\bibitem[{Marsch(2006)}]{marsch2006kinetic}
Marsch, E. 2006, Living Reviews in Solar Physics, 3, 1

\bibitem[{Mozer {et~al.}(2020{\natexlab{a}})Mozer, Agapitov, Bale, Bonnell,
  Bowen, \& Vasko}]{mozer2020dc}
Mozer, F., Agapitov, O., Bale, S., {et~al.} 2020{\natexlab{a}}, Journal of
  Geophysical Research: Space Physics, 125, e2020JA027980

\bibitem[{Mozer {et~al.}(2020{\natexlab{b}})Mozer, Bonnell, Bowen, Schumm, \&
  Vasko}]{mozer2020large}
Mozer, F., Bonnell, J., Bowen, T., Schumm, G., \& Vasko, I. 2020{\natexlab{b}},
  The Astrophysical Journal, 901, 107

\bibitem[{Mozer \& Chen(2013)}]{mozer2013parallel}
Mozer, F., \& Chen, C. 2013, The Astrophysical Journal Letters, 768, L10

\bibitem[{Narita(2018)}]{narita2018space}
Narita, Y. 2018, Living reviews in solar physics, 15, 1

\bibitem[{Narita {et~al.}(2016)Narita, Nakamura, Baumjohann, Glassmeier,
  Motschmann, Giles, Magnes, Fischer, Torbert, Russell,
  {et~al.}}]{narita2016electron}
Narita, Y., Nakamura, R., Baumjohann, W., {et~al.} 2016, The Astrophysical
  Journal Letters, 827, L8

\bibitem[{Podesta(2009)}]{podesta2009dependence}
Podesta, J. 2009, The Astrophysical Journal, 698, 986

\bibitem[{Ruan {et~al.}(2016)Ruan, He, Zhang, Vocks, Marsch, Tu, Peter, \&
  Wang}]{ruan2016kinetic}
Ruan, W., He, J., Zhang, L., {et~al.} 2016, The Astrophysical Journal, 825, 58

\bibitem[{Sahraoui {et~al.}(2009)Sahraoui, Goldstein, Robert, \&
  Khotyaintsev}]{sahraoui2009evidence}
Sahraoui, F., Goldstein, M., Robert, P., \& Khotyaintsev, Y.~V. 2009, Physical
  review letters, 102, 231102

\bibitem[{Salem {et~al.}(2012)Salem, Howes, Sundkvist, Bale, Chaston, Chen, \&
  Mozer}]{salem2012identification}
Salem, C.~S., Howes, G., Sundkvist, D., {et~al.} 2012, The Astrophysical
  Journal Letters, 745, L9

\bibitem[{Santol{\'\i}k {et~al.}(2003)Santol{\'\i}k, Parrot, \&
  Lefeuvre}]{santolik2003singular}
Santol{\'\i}k, O., Parrot, M., \& Lefeuvre, F. 2003, Radio Science, 38

\bibitem[{Shi {et~al.}(2021)Shi, Zhao, Huang, Wang, Wu, Chen, Hu, Kasper, \&
  Bale}]{shi2021parker}
Shi, C., Zhao, J., Huang, J., {et~al.} 2021, The Astrophysical Journal Letters,
  908, L19

\bibitem[{Sonnerup \& Cahill~Jr(1967)}]{sonnerup1967magnetopause}
Sonnerup, B.~{\"O}., \& Cahill~Jr, L. 1967, Journal of Geophysical Research,
  72, 171

\bibitem[{Stansby {et~al.}(2016)Stansby, Horbury, Chen, \&
  Matteini}]{stansby2016experimental}
Stansby, D., Horbury, T., Chen, C., \& Matteini, L. 2016, The Astrophysical
  Journal Letters, 829, L16

\bibitem[{Stix(1992)}]{stix1992waves}
Stix, T.~H. 1992, Waves in plasmas (Springer Science \& Business Media)

\bibitem[{Sun {et~al.}(2020)Sun, Zhao, Liu, Xie, \& Wu}]{sun2020electron}
Sun, H., Zhao, J., Liu, W., Xie, H., \& Wu, D. 2020, The Astrophysical Journal,
  902, 59

\bibitem[{Swanson(2003)}]{swanson2003plasma}
Swanson, D.~G. 2003, Plasma waves (CRC Press)

\bibitem[{Tong {et~al.}(2019)Tong, Vasko, Artemyev, Bale, \&
  Mozer}]{tong2019statistical}
Tong, Y., Vasko, I.~Y., Artemyev, A.~V., Bale, S.~D., \& Mozer, F.~S. 2019, The
  Astrophysical Journal, 878, 41

\bibitem[{Vech {et~al.}(2020)Vech, Martinovic, Klein, Malaspina, Bowen,
  Verniero, Paulson, Dudok, de~Wit, Huang, {et~al.}}]{vech2020wave}
Vech, D., Martinovic, M.~M., Klein, K.~G., {et~al.} 2020, arXiv preprint
  arXiv:2010.15189

\bibitem[{Verniero {et~al.}(2020)Verniero, Larson, Livi, Rahmati, McManus,
  Pyakurel, Klein, Bowen, Bonnell, Alterman, {et~al.}}]{verniero2020parker}
Verniero, J., Larson, D., Livi, R., {et~al.} 2020, The Astrophysical Journal
  Supplement Series, 248, 5

\bibitem[{Verscharen {et~al.}(2013)Verscharen, Bourouaine, \&
  Chandran}]{verscharen2013instabilities}
Verscharen, D., Bourouaine, S., \& Chandran, B.~D. 2013, The Astrophysical
  Journal, 773, 163

\bibitem[{Verscharen {et~al.}(2019)Verscharen, Klein, \&
  Maruca}]{verscharen2019multi}
Verscharen, D., Klein, K.~G., \& Maruca, B.~A. 2019, Living reviews in solar
  physics, 16, 1

\bibitem[{Wicks {et~al.}(2016)Wicks, Alexander, Stevens, Wilson~III, Moya,
  Vi{\~n}as, Jian, Roberts, O’Modhrain, Gilbert, {et~al.}}]{wicks2016proton}
Wicks, R., Alexander, R., Stevens, M., {et~al.} 2016, The Astrophysical
  Journal, 819, 6

\bibitem[{Woodham {et~al.}(2019)Woodham, Wicks, Verscharen, Owen, Maruca, \&
  Alterman}]{woodham2019parallel}
Woodham, L.~D., Wicks, R.~T., Verscharen, D., {et~al.} 2019, The Astrophysical
  Journal Letters, 884, L53

\bibitem[{Yang {et~al.}(2017)Yang, Zhang, He, Tu, Li, Wang, \&
  Wang}]{yang2017formation}
Yang, L., Zhang, L., He, J., {et~al.} 2017, The Astrophysical Journal, 851, 121

\bibitem[{Yoon(2017)}]{yoon2017kinetic}
Yoon, P.~H. 2017, Reviews of Modern Plasma Physics, 1, 1

\bibitem[{Zank {et~al.}(2020)Zank, Nakanotani, Zhao, Adhikari, \&
  Kasper}]{zank2020origin}
Zank, G., Nakanotani, M., Zhao, L.-L., Adhikari, L., \& Kasper, J. 2020, The
  Astrophysical Journal, 903, 1

\bibitem[{Zhang {et~al.}(2012)Zhang, Feng, Chen, Xu, Li, \&
  Forrest}]{zhang20122009}
Zhang, Y., Feng, W., Chen, Y., {et~al.} 2012, Geophysical Journal
  International, 191, 1417

\bibitem[{Zhao {et~al.}(2019)Zhao, Feng, Wu, Pi, \& Huang}]{zhao2019generation}
Zhao, G., Feng, H., Wu, D., Pi, G., \& Huang, J. 2019, The Astrophysical
  Journal, 871, 175

\bibitem[{Zhao {et~al.}(2020)Zhao, Lin, Wang, Wu, Feng, Liu, Zhao, \&
  Li}]{zhao2020observational}
Zhao, G.~Q., Lin, Y., Wang, X., {et~al.} 2020, Geophysical Research Letters,
  47, e2020GL089720

\bibitem[{Zhao {et~al.}(2018)Zhao, Wang, Dunlop, He, Dong, Wu, Khotyaintsev,
  Ergun, Russell, Giles, {et~al.}}]{zhao2018modulation}
Zhao, J., Wang, T., Dunlop, M., {et~al.} 2018, The Astrophysical Journal, 867,
  58

\bibitem[{Zhao {et~al.}(2021)Zhao, Zank, He, Telloni, Hu, Li, Nakanotani,
  Adhikari, Kilpua, Horbury, {et~al.}}]{zhao2021turbulence}
Zhao, L., Zank, G., He, J., {et~al.} 2021, arXiv preprint arXiv:2102.03301

\bibitem[{Zhu {et~al.}(2019)Zhu, He, Verscharen, \& Zhao}]{zhu2019composition}
Zhu, X., He, J., Verscharen, D., \& Zhao, J. 2019, The Astrophysical Journal,
  878, 48

\end{thebibliography}
\bibliographystyle{aasjournal}


\end{document}